\documentclass{article}

\usepackage{PRIMEarxiv}

\usepackage[utf8]{inputenc} 
\usepackage[T1]{fontenc}    
\usepackage{hyperref}       
\usepackage{url}            
\usepackage{booktabs}       
\usepackage{amsfonts}       
\usepackage{nicefrac}       
\usepackage{microtype}      
\usepackage{lipsum}
\usepackage{fancyhdr}       
\usepackage{graphicx}       
\graphicspath{{media/}}     
\usepackage{graphicx}%
\usepackage{multirow}%
\usepackage{amsmath,amssymb,amsfonts}%
\usepackage{amsthm}%
\usepackage{mathrsfs}%
\usepackage[title]{appendix}%
\usepackage{xcolor}%
\usepackage{textcomp}%
\usepackage{manyfoot}%
\usepackage{booktabs}%
\usepackage{algorithm}%
\usepackage{algorithmicx}%
\usepackage{algpseudocode}%
\usepackage{listings}%
\usepackage{tabularx}%
\usepackage{array}%
\usepackage{multirow}%
\usepackage{subcaption}

\usepackage{colortbl} 
\title{Two-Fold Byzantine Fault Tolerance Algorithm:
Byzantine Consensus in Blockchain

}

\author{
  Mohammad R. Shakournia \\
  School of Electrical and Computer Engineering \\
  University of Tehran \\
  Tehran,Iran\\
  \texttt{shakournia@ut.ac.ir} \\
   \And
  Pooya Jamshidi \\
  School of Electrical and Computer Engineering \\
  University of Tehran \\
  Tehran,Iran\\
  \texttt{pooya.jamshidi@ut.ac.ir} \\
    \And
  Hamid Reza Faragardi \\
  Research Institute of Sweden \\
  Stockholm,Sweden \\
  \texttt{hrfa@kth.se} \\
    \And
  Nasser yazdani \\
  School of Electrical and Computer Engineering\\
  University of Tehran\\
  Tehran,Iran\\
  \texttt{yazdani@ut.ac.ir}\\
}

\begin{document}
\maketitle

\begin{abstract}
Blockchain technology offers a decentralized and secure method for storing and authenticating data, rendering it well-suited for various applications such as digital currencies, supply chain management, and voting systems. However, the decentralized nature of blockchain also exposes it to vulnerabilities, particularly Byzantine faults, which arise when nodes in the network behave maliciously or encounter unexpected failures. Such incidents can result in inconsistencies within the blockchain and, in extreme scenarios, lead to a breakdown in consensus. Byzantine fault-tolerant consensus algorithms are crafted to tackle this challenge by ensuring that network nodes can agree on the blockchain's state even in the presence of faulty or malicious nodes. To bolster the system's resilience against these faults, it is imperative to detect them within the system. However, our examination of existing literature reveals a prevalent assumption: solutions typically operate under constraints regarding the number of faulty nodes. Such constraints confine the proposed solutions to ideal environments, limiting their practical applicability. In response, we propose a novel approach inspired by social paradigms, employing a trusted and fully monitored communication sub-process to detect Byzantine nodes. Upon detection, these nodes can be either disregarded in the consensus-building process, subjected to penalties, or undergo modifications as per the system's policy. Finally, we statistically demonstrate that our approach achieves a detection probability that exceeds 95\% for Byzantine nodes. In essence, our methodology ensures that if Byzantine nodes exhibit malicious behavior, healthy nodes can identify them with a confidence level of 95\%.
\end{abstract}


\section{Introduction}
The blockchain serves as a decentralized distributed ledger system that securely and transparently records transactions. Performing on a peer-to-peer network, the nodes engage in communication to maintain the integrity of the ledger. Using cryptographic hash functions, the blockchain generates digital signatures to authenticate transactions and thwart tampering attempts. Message passing assumes a pivotal role in blockchain operations, enabling nodes to exchange information and validate transactions. When a user initiates a transaction, it is distributed across all nodes of the network. Each node then verifies the transaction's validity through a consensus algorithm. Upon confirmation, the transaction is appended to the blockchain, making it immutable. Failure to reach a consensus within the network can precipitate blockchain failures. Disagreements among nodes on the validity of the transaction or the state of the blockchain can result in forks. Forks occur when multiple blocks are generated concurrently, leading to network division as nodes accept divergent blockchain versions. This discord impedes consensus on valid transactions, complicating or obstructing node agreement on network state. 

The lack of consensus arises when the nodes within a distributed system cannot agree on its current state. This situation can arise from various factors, including network failures, delays, or Byzantine faults. Byzantine faults, in particular, play a pivotal role in causing consensus failures within distributed systems like Blockchain. A Byzantine fault occurs when nodes in the network behave unexpectedly or maliciously, such as by transmitting false or deceptive data or by ignoring requests. These kinds of fault can result in nodes holding divergent perspectives on the system's state, ultimately leading to a breakdown in consensus. Blockchain stands out as one of the most vulnerable distributed systems susceptible to consensus failures.

 In blockchain, addressing the lack of consensus is achievable through consensus mechanisms or directly via consensus algorithms, also known as consensus protocols. Proof of Work (PoW) \cite{Nakamoto2008}, Proof of Stake (PoS) \cite{King2012}, Practical Byzantine Fault Tolerance (PBFT) \cite{Castro1999}, and Delegated Proof of Stake (DPoS) \cite{larimer2017dpos} represent common consensus mechanisms employed in blockchain networks, leveraging economic incentives and computational puzzles to achieve consensus. Ripple Protocol Consensus Algorithm (RPCA) \cite{Schwartz2014} and Stellar Consensus Protocol (SCP) \cite{Mazieres2015}, on the other hand, are consensus algorithms prioritizing fault tolerance and consensus among nodes, even amidst adversarial scenarios. These algorithms are specifically crafted to ensure that network nodes converge on the system's state, even when faced with malicious actors or network failures. Consensus algorithms enable blockchain to function as a trustless and decentralized system, thereby furnishing users with security and transparency.

Consensus mechanisms like PoW and PoS contribute to fault tolerance by necessitating a majority agreement among nodes regarding the validity of transactions and blocks prior to their inclusion in the blockchain. RPCA employs a federated consensus model, where a limited number of trusted nodes are tasked with upholding network consensus. DPoS presents a more efficient and scalable alternative to PoW and PoS. In DPoS, token holders elect a set number of delegates to authenticate transactions and generate new blocks on the blockchain. SCP enables nodes to designate a subset of other nodes they trust, safeguarding against disruptive actions by malicious nodes through quorum slices and threshold encryption. PBFT confirms a transaction only when at least 66.6\% of nodes in the network concur on the ledger's state. Section 3 offers further elucidation on these algorithms. Building upon the classical Byzantine Generals Problem \cite{Leslie1998}, Lamport et al. demonstrated that a few traitorous nodes cannot lead loyal nodes to an incorrect decision. In their model, there are 33\% traitorous nodes within the network. They concluded that an algorithm ensuring consensus will function under the following conditions: a) all loyal nodes eventually converge on the same decision, and b) a few traitorous nodes cannot influence loyal nodes to make an erroneous decision \cite{Leslie1998}.

It has been suggested in various articles that certain assumptions underlying these algorithms, such as a trusted group of nodes or a fixed upper bound on the number of nodes, must be met for the algorithm to function effectively. Even though some of these algorithms have been deployed in practical environments, we must consider: how can we limit the number of faulty nodes in the system? Therefore, maintaining such conditions becomes challenging. Consequently, there is a need for a consensus algorithm that can accommodate varying numbers of faulty nodes without being constrained by the number of Byzantine nodes. This is our focus—addressing the issue of the number of faulty nodes in the system, which is inherently unpredictable.

We introduce the TDBA algorithm to tackle this challenge. TDBA is a two-part protocol designed to empower all processes within the Blockchain system to identify Byzantine nodes. Upon detection, non-faulty nodes will disregard messages from these identified Byzantine nodes. Furthermore, nodes will collaborate to maintain a blacklist of faulty nodes, ensuring that non-faulty nodes maintain a comprehensive list of identified problematic nodes. Importantly, TDBA operates without the need for any federation or pre-trusted group of nodes; all nodes possess the capability to independently detect faulty nodes.

The rest of this paper is organized as follows; Section 2 surveys the related work. In Sections 3 and 4, we present the proposed solutions and implementation. Section 5 illustrates the simulation results. Section 6 concludes this paper .Finally, Section 7 discuss the future works.

\section{Related Work}
Consensus protocols have been extensively studied in the field of distributed systems, with various approaches proposed over the years, as summarized in \hyperref[table1]{Table 1}. These approaches generally fall into two categories: leader-based and leaderless. This section will explore both types. Additionally, Shaohua Wan et al. \cite{wan2020recent} have classified consensus protocols into permissioned and permissionless categories based on access levels. Although our primary concern is the number of faulty nodes a consensus protocol can tolerate rather than access levels, we will operate under the assumption that our environment is permissionless. This means any node can join and participate in the consensus process.

One of the earliest and most fundamental algorithms in distributed systems, introduced by Leslie Lamport in 1990, is called Paxos \cite{Leslie1998,Castro1999}. It is a leader-based distributed consensus algorithm designed to achieve fault-tolerant consensus. In Paxos, a single designated leader, known as the "proposer," is responsible for proposing values that a group of participants, termed "acceptors," must agree upon. Through a series of message exchanges, the proposer attempts to obtain a majority of acceptors' agreement on a proposed value. The Paxos system ensures fault tolerance by tolerating up to one third of faulty nodes (33\%). 

The most widely known of these protocols is Practical Byzantine Fault Tolerance (PBFT), proposed by Castro and Liskov in 1999 \cite{Castro1999}. PBFT is a leader-based protocol that can tolerate up to one-third (33\%) of the nodes in the network being faulty or malicious, making it a popular choice for blockchain systems. However, PBFT requires a high degree of network communication, which can limit its scalability.

In PBFT, nodes must wait for the leader to propose new transactions before they can validate them. This can increased latency, particularly in networks with high packet loss rates or significant message propagation delays. Zyzzyva \cite{Kotla2008} was developed in 2008. Like PBFT, Zyzzyva elects a leader from among the nodes in the system to propose new transactions and order them. The other nodes in the system then validate the leader's proposals and reach consensus on their order. One of the key differences between Zyzzyva and other BFT protocols, such as PBFT, is that Zyzzyva employs speculative execution to improve performance. Specifically, Zyzzyva allows non-leader nodes to execute requests speculatively, without waiting for the leader's proposal. This reduce the overall system latency, as nodes do not need to wait for the leader to propose a new transaction before they starting validation. Zyzzyva achieves a fault tolerance of up to one-third of the total number of nodes in the network.

After Kotla et al. introduced Zyzzyva, Nakamoto published a pioneering article that, while not specifically focused on Byzantine Fault Tolerance (BFT), became an essential part of blockchain technology. This article introduced the concept of Proof of Work (PoW) \cite{Nakamoto2008}, a leaderless approach that allows participants to compete in solving computationally intensive puzzles to validate and add new blocks to the blockchain. PoW is the first and most widely adopted consensus protocol in current blockchain systems. Essentially, PoW confirms that a certain amount of work has been completed. For example, before sending a mail message, proof of work must be provided. A PoW network can tolerate faults in up to one-half (50\%) of its nodes without compromising security and integrity.

Since complex mathematical problems requires significant computational power, PoW is not energy efficient. This limitation has led to many variations in PoW. The first variation, Proof of Stack (PoS)\cite{King2012}, was proposed in 2012. PoS is a consensus mechanism that operates based on participants' ownership or stake in a cryptocurrency. PoS selects validators to add new blocks to the blockchain based on the amount of cryptocurrency they hold and are willing to "stake" to participate in the consensus process. Validators, chosen randomly or through a voting process depending on the specific implementation, do not need to compete to solve complex problems, significantly reducing the energy required to participate in the consensus process. PoS suffers from the "nothing-at-stake" problem, meaning validators have nothing to lose when a malicious actor gains control of a significant portion of the network. This is possible because they can create multiple nodes without substantial investment, compromising its security and reliability. In 2013, Proof of Capacity (PoC) \cite{Park2015} was proposed as a leaderless consensus mechanism where participants prove their storage capacity to validate new blocks. In PoC, participants pre-allocate a certain amount of storage to mine and contribute to the consensus process. The mining power is determined by available storage space, rather than computational power or token ownership. Participants generate plots or cryptographic proofs to demonstrate their storage capacity and compete to solve complex puzzles related to the allocated space. Proof of Space-Time (PoST) \cite{Park2015}, also proposed in 2013, utilizes storage capacity like PoC but additionally considers the time taken to compute the proof of space. Although PoC and PoST address energy efficiency issues, they also have their own limitations. For example, PoC requires significant hard drive space, making it challenging for smaller miners to compete with larger ones. This problem allows larger miners with more resources to dominate the network. As a solution to this limitation, a new variant called Proof of Burn (PoB) \cite{P4Titan2014} was introduced in 2014, where miners "burn" or destroy a certain amount of cryptocurrency tokens to prove their commitment to the network. This process involves sending tokens to an unspendable address, removing them from circulation forever. Although these variations of PoW were not specifically designed for Byzantine faults, in 2014, Delegated Proof of Stake (DPoS) \cite{larimer2017dpos}, was proposed as a solution to this problem. DPoS introduces a delegated governance model where token holders vote to select a limited number of trusted delegates who are responsible for validating transactions and creating blocks. These delegates take turns producing blocks in a round-robin fashion. The limitation of DPoS is that it can become centralized when a few powerful delegates accumulate significant voting power. This centralization increases the risk of collusion among delegates, which can compromise both the security and decentralization of the network, challenging the foundational principles of blockchain technology. Proof of Elapsed Time (PoET) \cite{chen2017security} proposed in 2016 , leverages a trusted execution environment (TEE) and a randomized leader selection process. In PoET, each participant requests a random wait time from the TEE, and the participant with the shortest wait time becomes the leader for a specific block. The leader then proposes the next block, and other participants validate and endorse it. Unlike Delegated Proof of Stake (DPoS), PoET does not rely on a fixed set of delegates. Instead, the leader selection process is randomized, making it a leader-less consensus mechanism. A simpler approach for consensus in private or permissioned blockchain networks was proposed by Parity Technologies in 2017 called Proof of Authority (PoA) \cite{de2018pbft}. Unlike PoET, which uses a TEE to generate random waiting times, PoA relies on the reputation and identity of pre-selected validators to ensure the security and validity of the network. PoA, like the other protocols mentioned, can continue to function effectively even if up to half of the nodes are faulty or malicious. Each of the variations of PoW, except DPoS, is not inherently Byzantine Fault Tolerant (BFT). Generally, they all have a fault tolerance of 50\%, meaning the protocol can continue to function even if up to half of the nodes become faulty. 

Over the past decade, Byzantine Faults have posed significant challenges in blockchain technology. In response, the Ripple Protocol Consensus Algorithm (RPCA) \cite{Schwartz2014} was introduced in 2014 as a successor to BFT protocols like Zyzzyva and DPoS. RPCA, fundamentally different from Zyzzyva, uses a consensus mechanism that enables the Ripple network to achieve agreement on the states of its distributed ledger. Unlike Zyzzyva’s reliance on Byzantine fault tolerance techniques, RPCA utilizes a consensus algorithm based on a unique node list (UNL). In RPCA, each participant keeps a list of trusted nodes, their UNL, which is crucial for validating and propagating transactions. Participants engage in continuous communication with their UNL nodes to exchange and verify ledger versions. Through iterative voting rounds, they reach a consensus by identifying the ledger version that the majority in their UNL supports. As a leader-based consensus protocol, RPCA can tolerate up to 20\% faulty nodes.

Tendermint, a protocol specifically designed for blockchain, was developed by Buchman \cite{buchman2016tendermint}. As a leader-based Byzantine fault-tolerant (BFT) consensus protocol, Tendermint employs a round-robin process to select a leader who proposes a new block of transactions in each round. Validators independently verify this block and cast votes on whether to accept it. Once a sufficient number of votes are obtained, the block is deemed pre-committed. This pre-committed block is then broadcasted and committed by all validators, solidifying its place in the blockchain. Tendermint ensures Byzantine fault tolerance by allowing up to one-third (33\%) of the validators to be malicious or arbitrary without compromising the network's integrity.

The Stellar Consensus Protocol (SCP) \cite{Mazieres2015}, a decentralized consensus algorithm, was developed specifically for the Stellar blockchain. Unlike Tendermint, SCP operates without a designated leader to propose blocks. Instead, it utilizes a federated voting process where nodes form a network of quorums—groups of nodes that intersect to ensure consensus on the order and validity of transactions. Each node participates in its respective quorum, with agreement reached through iterative voting rounds. SCP employs the Federated Byzantine Agreement (FBA) model, wherein each node trusts a specific set of other nodes known as "quorum slices." This trust graph structure facilitates efficient and decentralized consensus. SCP is designed to tolerate faults in up to one-third of the nodes, whether faulty or malicious.

In 2016, Miller et al. introduced HoneyBadgerBFT \cite{Miller2016}, a leaderless Byzantine fault-tolerant (BFT) consensus protocol designed for decentralized networks. Differing from leader-based protocols, HoneyBadgerBFT does not depend on a single leader to propose blocks or make decisions. It operates on a robust asynchronous network model that allows nodes to communicate and exchange messages without the need for strict timing assumptions. The protocol employs randomized secret sharing and cryptographic mixing to ensure Byzantine fault tolerance, enabling honest nodes to achieve consensus on the order and validity of transactions, even with up to one-third (33\%) of the nodes being malicious or faulty.

Introduced by Gilad et al. in 2017, Algorand \cite{gilad2017algorand} is a leaderless consensus protocol capable of tolerating up to 20\% of nodes failing. It employs a novel approach known as Pure Proof of Stake (PPoS), which uses cryptographic sortition to randomly select a committee of nodes for each round of the consensus process. Committee members then use a Verifiable Random Function (VRF) to propose and vote on blocks for each round. Algorand is designed to be highly scalable, secure, and energy-efficient, addressing key challenges in blockchain technology. 

While Algorand employs cryptographic sortition for consensus, Avalanche, proposed by Team Rocket in 2018 \cite{rocket2018snowflake}, uses a distinct approach called a metastable mechanism. This method is designed to be highly scalable and resilient against network partitions and malicious attacks. It operates by relying on a large number of nodes that independently and simultaneously assess the validity of transactions. Nodes continuously sample the opinions of others in the network until they collectively converge on a single decision. Like Algorand, Avalanche is a leaderless protocol, capable of tolerating up to one-third (33\%) of the nodes being faulty.

The Flexible Byzantine Fault Protocol (FBFT) \cite{malkhi2019flexible} was introduced in 2019 to address the heterogeneity of clients within a network, which can vary in their requirements for resilience and timing. FBFT maintains a separation of the protocol from the fault model, enabling diverse clients with different fault and timing assumptions (whether synchronous or not) to participate in the same protocol. It allows clients to operate with varying message delay bounds and commit at their own pace. Additionally, FBFT introduces a new type of fault, the alive-but-corrupt (a-b-c) fault, which compromises safety but not liveness. Incorporating these features enables FBFT to tolerate a combination of faults (Byzantine plus a-b-c) beyond the traditional resilience limits, such as the one-third threshold for Byzantine faults.

In 2019, Sheng Gao and colleagues introduced the T-PBFT consensus algorithm, as outlined in their paper \cite{gao2019t}. This algorithm enhances the performance and security of the existing Practical Byzantine Fault Tolerance (PBFT) consensus methods. T-PBFT incorporates an EigenTrust-based reputation system to boost fault tolerance. This system enables network nodes to assess the trustworthiness of others based on historical behavior. Nodes with higher reputation scores have greater influence in the consensus process, whereas those with lower scores have less influence. This strategy aims to protect the consensus process from malicious actors. Under this protocol, nodes with a trust value 'd' between 0 and 1 $0 < d \leq 1$ engage in the consensus process. If d equals 1, it implies all nodes participate. The resilience of the T-PBFT, dependent on the value of 'd', can handle up to $\left( 1 - \frac{2}{3} d \right) N - \frac{1}{3}$ failures.

Yu Zhan and his team have identified that the Practical Byzantine Fault Tolerance (PBFT) becomes less efficient as the number of nodes increases, due to its complexity. To address this issue, they introduced a new protocol named Delegated Randomization Byzantine Fault Tolerance (DRBFT), as described in their study \cite{zhan2021drbft}. This protocol aims to fairly select delegate nodes from all participants to implement the PBFT using a Random Selection (RS) algorithm. The DRBFT process involves three steps. Initially, all nodes vote to select a fixed number of candidates, with each node voting once per period. The top N vote-getters become candidates. A subset of these candidates is then selected through the RS algorithm to act as councillors, who execute the PBFT consensus protocol to generate new blocks. Although the exact fault tolerance isn't specified, the use of PBFT suggests that it can handle up to a third of the nodes being faulty.

Security is a major concern for blockchain networks, which must contend with various assumptions about network timing, such as synchronous and asynchronous conditions. These assumptions can undermine the universality and portability of the protocol. For instance, changes in network conditions could compromise the original guarantees of persistence or liveness. To address this, the Flexible BFT \cite{malkhi2019flexible} enhances the flexibility of the BFT protocol to adapt to different network environments. It restructures the BFT protocol and introduces Flexible Byzantine Quorums. Flexible BFT enables a ledger to support nodes with varying assumptions, including synchronous or asynchronous networks, and different proportions of Byzantine nodes. It requires clients to select appropriate commitment rules based on the network conditions. Adhering to a commitment rule designed for synchronous networks in an asynchronous environment can jeopardize security, as highlighted in reference \cite{Xu2023}. To overcome this challenge, Momose et al. \cite{Momose2021} developed a multi-threshold Byzantine Fault Tolerance (BFT) protocol that adapts to various network timings using a single commitment rule. This protocol can withstand up to two-thirds (66\%) of faults in a secure synchronous network and one-third (33\%) of faults in secure asynchronous or partially synchronous networks. 

In spite of recent improvements in BFT protocols, state-of-the-art designs still suffer from suboptimal performance. The Dumbo-NG \cite{gao2022dumbo} protocol was proposed in 2022 by Yingzi Gao et al. to address the issue of throughput and latency. Using the new protocol structure, transaction dissemination and asynchronous agreement can be carried out concurrently. In this way, the throughput-latency tension can be resolved, enabling peak throughput to be achieved with minimal latency increase, and the censorship threat can be defeated by agreeing to output every honest node's transactions. In this article, the authors consider that up to $f = \left\lfloor \frac{n - 1}{3} \right\rfloor$ nodes might be fully controlled by the adversary. As a follow-up to Dumbo-NG, Speeding Dumbo \cite{guo2022speeding} aims to provide an asynchronous BFT consensus protocol that is more practical and further improves the efficiency of Dumbo-based \cite{rocket2018snowflake} protocols which consists of two atomic broadcast protocols Dumbo1 and Dumbo2,takes another approach to improve HoneyBadgerBFT. Despite all this studies, fault tolerance is still on one-third of faulty nodes and has not improved.

In \hyperref[table1]{Table 1}, we present a comparative overview of the complexity orders of various protocols relative to our own. We first examined time complexity and found that some protocols operate at a constant time complexity, or O(1), indicating that they do not inherently scale with the size of the network or transaction volume. For example, the time complexity of Proof of Work (PoW) is probabilistic; typically, finding a solution takes a predetermined time, such as 10 minutes for Bitcoin, although the time for individual attempts can vary widely. This variability also applies to other protocols with O(1) time complexity, prompting us to include an 'Execute Time' column in the table for further clarification. This column has two entries: 'High' for execution times on the order of minutes, and 'Low' for execution times on the order of seconds. While our algorithm may outpace those involving more intensive computations, the Zyzzyva protocol is designed for efficiency under any implementation or network condition. We classify Zyzzyva’s complexity as O(t), acknowledging that it does not adhere to a fixed number of rounds like some other BFT consensus algorithms, which contributes to its less predictable time complexity. Furthermore, we observed that our algorithm matches or surpasses other protocols in terms of space complexity. 

\begin{table}[ht]
    \centering
    \resizebox{\textwidth}{!}{ 
    {\Large
    \begin{tabular}{lccccccccc}
        \toprule
        Protocol & Leader & Time Complexity & Space Complexity & Message Complexity & Scalability & Fault Tolerance (\%) & BFT & Execution Time \\
        \midrule
        PoW (2008) & No & O(1) & O(1) & O(t) & High & 50\% & No & High \\  
        PoST (2013) & No & O(1) & O(n) & O(t) & High & 50\% & No & High \\  
        PoC (2013) & No & O(1) & O(h) & O(t) & High & 50\% & No & High \\  
        PoB (2014) & No & O(1) & O(h) & O(t) & High & 50\% & No & High \\  
        PoET (2016) & No & O(1) & O(1) & O(t) & High & 33\% & No & High \\  
        HoneyBadgerBFT (2016) & No & O(n²) & O(n) & O(n³) & High & 33\% & Yes & High \\  
        Algorand (2017) & No & O(1) & O(n) & O(t) & High & 20\% & Yes & High \\  
        Avalanche (2018) & No & O(1) & O(t) & O(nS) & High & 33\% & Yes & Low \\  
        TDBA (2023)* & No & O(n) & O(n) & O(n²) & High & 95\% & Yes & Low \\  
        Paxos (1989) & Yes & O(n²) & O(n) & O(n²) & High & 33\% & No & Low \\  
        PBFT (1999) & Yes & O(n²) & O(n²) & O(n²) & Low & 33\% & Yes & Low \\  
        Zyzzyva (2008) & Yes & O(t) & O(n) & O(n) & High & 33\% & Yes & Low \\  
        PoS (2012) & Yes & O(1) & O(n) & O(t) & High & 50\% & No & High \\  
        DpoS (2014) & Yes & O(1) & O(n) & O(t) & High & 50\% & Yes & High \\  
        Ripple (RPCA 2014) & Yes & O(1) & O(n) & O(t) & High & 20\% & Yes & Low \\  
        Tendermint (2014) & Yes & O(1) & O(n) & O(t) & High & 33\% & Yes & High \\  
        FBA (2015) & Yes & O(1) & O(n) & O(t) & High & 33\% & Yes & High \\  
        PoA (2017) & Yes & O(1) & O(n) & O(t) & High & 50\% & No & Low \\  
        MT-BFT (2021) & Yes & O(n²) & O(n²) & O(n²) & Low & 66\% & Yes & Low \\  
        \bottomrule
        \label{table1}
    \end{tabular}
    }}
    \caption{Consensus Protocol Comparison}
\end{table}

Although some protocols operate with a constant time complexity, or O(1), the primary resource they consume, like in Proof of Work (PoW), is computing power (CPU/GPU/ASIC) rather than storage. However, space complexity is significant in Proof of Capacity (PoCs), where participants need considerable storage to create and maintain data plots. Proof of Elapsed Time (PoET) requires secure hardware or a trusted execution environment to generate random wait times and ensure fairness; beyond this, PoET does not demand much storage space. Avalanche, on the other hand, records a list of transactions and their preferences, resulting in space complexity that grows with the number of transactions. The message complexity column in our table indicates the volume of messages exchanged between nodes to achieve consensus. While our method may seem somewhat complex compared to others, it's crucial to understand that this complexity allows us to overcome limitations seen in other protocols. Not only does our approach enable us to identify Byzantine nodes, but it also supports achieving reliable consensus even if over 50\% of the nodes act maliciously. Message complexity is categorized as O(t), indicating that the number of messages depends on the blocklist size and transaction throughput, but remains generally manageable.

The meaning of scalability is that with the scaling of the number of nodes, the performance of the algorithm does not decrease. However, in the case of TDBA, despite doubling the messages, the message complexity order does not increase, thereby ensuring scalability with a complexity of O(n\textsuperscript{2}).

In the following section, we introduce our proposed method, TDBA. According to Section II, the current algorithms in the literature offer up to 66\% fault tolerance. However, our research seeks to enhance this capacity. We developed TDBA based on the existing consensus protocol literature, with the goal of enhancing the scalability and efficiency of consensus mechanisms in these systems, while also preserving a high level of security and fault tolerance.

\section{TWO-FOLD BFT}
\subsection{Prelimineirs}
\textbf{Reliable Broadcast (RBC). }Reliable broadcast is a communication protocol used in distributed systems to ensure that a message sent by one node is received by all other nodes in the system in a reliable and consistent manner. Reliable Broadcast (RBC) has the following properties:
\begin{itemize}

    \item \textbf{Validity:} If a correct process broadcasts a message, then all correct processes eventually deliver the message.
    \item  \textbf{Integrity:} If a process delivers a message \textit{m}, then \textit{m} was previously broadcast by some process.
    \item  \textbf{Agreement:} If a correct process delivers a message \textit{m}, then all correct processes eventually deliver \textit{m}, and deliver \textit{m} in the same order.
    \item  \textbf{No duplication:} No message is delivered more than once.

\end{itemize}

The algorithm uses RBC to broadcast transactions and blocks to all nodes in the network, to guarantee that messages are delivered reliably, even if some nodes fail, and to agree on which order transactions and blocks are added.

\textbf{Consensus criteria.} Consensus criteria in blockchain refer to the properties that a consensus algorithm must satisfy to ensure the reliability and security of the blockchain network.

\begin{itemize}
    \item  \textbf{Agreement:} All correct nodes in the network should eventually agree on the same value. This means that, at the end of the consensus process, all honest nodes should reach consensus on a single value or outcome.
    \item  \textbf{Validity:} The agreed-upon value should be a valid value proposed by one of the nodes in the system. In the context of blockchain, this means that only valid transactions can be included in the agreed-upon blocks.
    \item \textbf{Termination:} The consensus algorithm should eventually terminate, meaning that all correct nodes should reach a decision within a finite number of steps or rounds. This ensures that the consensus process does not run indefinitely.
    \item \textbf{Integrity:} If a correct node decides on a certain value, then that value must have been proposed by some correct node. This property prevents any unauthorized or malicious values from being agreed upon.
\end{itemize}

\textbf{Asynchronous Blockchain}. The timing of this system is Asynchronous. Unlike real-time blockchains, asynchronous blockchains are designed to run autonomously without synchronization between nodes.

\textbf{Fault model.} Our model assumes Byzantine faults with arbitrary behavior. We present protocols that tolerate adaptive corruption, which may occur anytime during protocol execution. It is said that a replica that executes a protocol faithfully and without errors is honest.
\subsection{Network Architecture}
In our study, Blockchain networks are peer-to-peer (P2P) and permissionless networks in which nodes communicate directly without relying on a central authority. By using Reliable Broadcast (RBC), all correct nodes receive the same set of messages in the same order. We assume an asynchronous communication model in which messages are delayed and processed indefinitely. A decent level of security is provided for messages by using digital signatures. An inbox is considered for each node, where all incoming messages are stored. After processing these messages, each node decides which value or transaction to add to its copy of the blockchain ledger. Nodes in the network have no distinguishing identity that can be used to discern Byzantine or correct nodes. Detection relies solely on their behavior and message content. Specifically, we focus on comparing messages received in nodes' inboxes with messages received in their child nodes. Each node creates its own child nodes for detecting Byzantine nodes, a process which will be explained in detail in the following section.

\begin{figure}[ht]
    \centering
    \begin{minipage}{0.43\textwidth}
        \centering
        \includegraphics[width=0.75\linewidth]{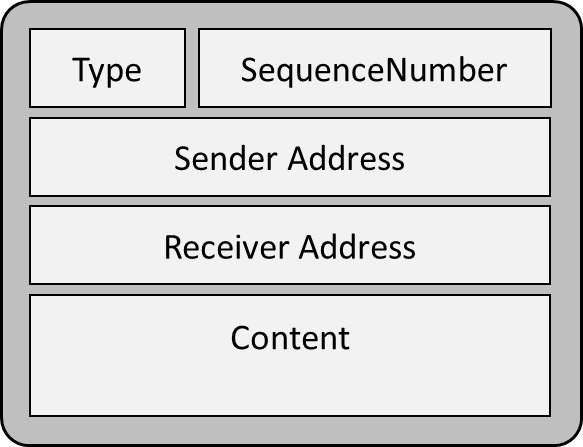}
        \caption{Message Structure}
        \label{fig1}
    \end{minipage}
    \hfill
    \begin{minipage}{0.45\textwidth}
        \centering
        \includegraphics[width=1\textwidth]{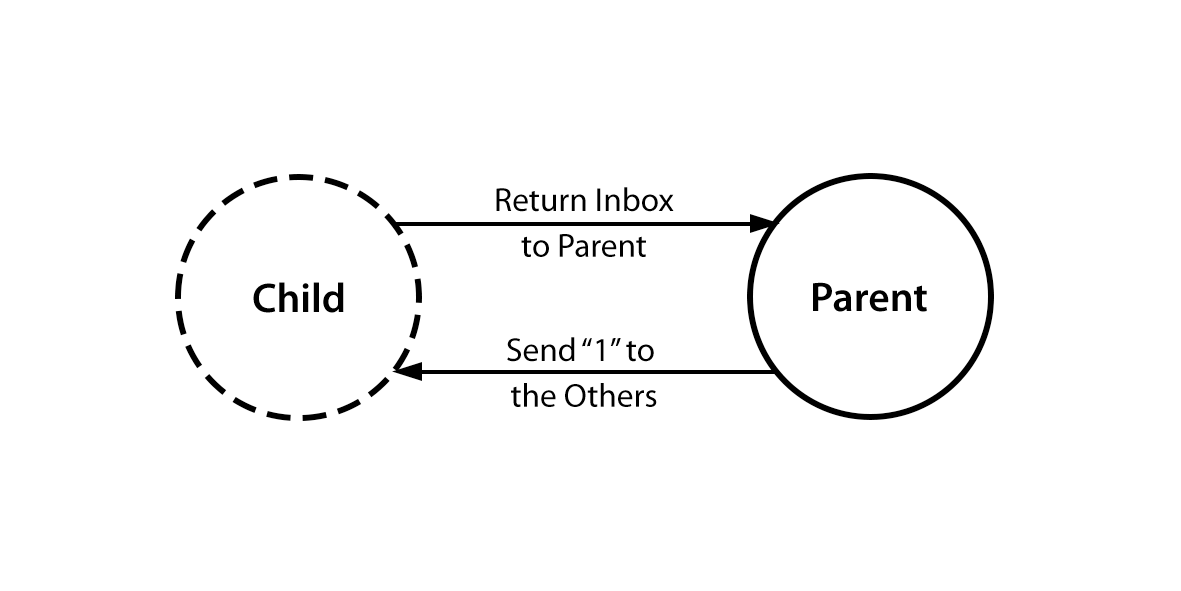}
        \caption{Communication between Parent and Child}
        \label{fig2}
    \end{minipage}
\end{figure}

Messages are crucial to our protocol, including key elements such as type, SequenceNumber, content, and sender/receiver addresses as shown in \hyperref[fig1]{Figure 1}. We have three message types: vote, transaction, and blacklist. Blacklist messages contain lists of Byzantine nodes, voting messages record consensus decisions, and transaction messages detail transaction data including the nonce. The SequenceNumber, assigned by the sender, ensures messages are processed in the order they were sent, maintaining reliable event sequencing without external synchronization.

\subsection{The Idea Behind TWO-FOLD BFT}
Each node uses an inbox to store incoming messages. However, with a Byzantine node present, the inboxes can vary across different nodes, potentially leading to consensus failures. Typically, healthy nodes receive a range of messages from a sender identified as Byzantine and are required to make decisions based on these. A major challenge is that nodes cannot converge on a common decision. The presence of Byzantine nodes doesn't cause the system to fail suddenly; rather, it leads to a gradual deterioration. This slow degradation is challenging for system administrators to detect and counteract. Although all nodes are aware of the existence of a problematic Byzantine node, pinpointing the specific one remains elusive.

To detect Byzantine nodes, we employ a difference detector to identify malicious messages. In environments prone to Byzantine faults, nodes lack shared information. Messages from a Byzantine node, although expected to be consistent, often differ, leading receivers to receive inconsistent messages from the same sender. In such systems, trust between nodes is absent. If a node identifies a Byzantine sender, it ignores these messages and warns others by sending a blacklist message that includes all identified Byzantine nodes.

Our proposed solution draws inspiration from social interactions where individuals often manage multiple email and social media accounts. Take Nick, a 33-year-old front-end developer from Wisconsin, who uses two email accounts: nickdeveloper@gmail.com and wisconsin\_frontend@yahoo.com. Nick is part of an email group where messages can be sent privately or publicly. An adversary unaware that both addresses belong to Nick might send different messages to each account. However, Nick can spot these discrepancies because he has access to both accounts. This unique position allows him to identify and address the adversary effectively.

To detect unequal messages, we propose a method similar to Nick’s, designed to handle Byzantine behavior which can disrupt consensus through a variety of messages. Our approach involves assigning each process a subprocess, or child process, that the parent process will closely monitor, as depicted in \hyperref[fig2]{Figure 2}. Each child process has its own identity, like an IP address. Because these Byzantine nodes are unaware that the child and parent processes are linked, they operate freely, sending messages as they wish without attempting to appear non-Byzantine. A child process broadcasts the same message as its parent and forwards any received message back to the parent. If the parent process notices discrepancies between the messages it sent and those received by its child, it identifies the sender as Byzantine and adds it to a blacklist, ignoring any further messages from blacklisted senders. As long as there are no Byzantine messages in the consensus process, it will function correctly. We describe our algorithm in the following section.
\begin{figure}
    \centering
    \includegraphics[width=1\linewidth]{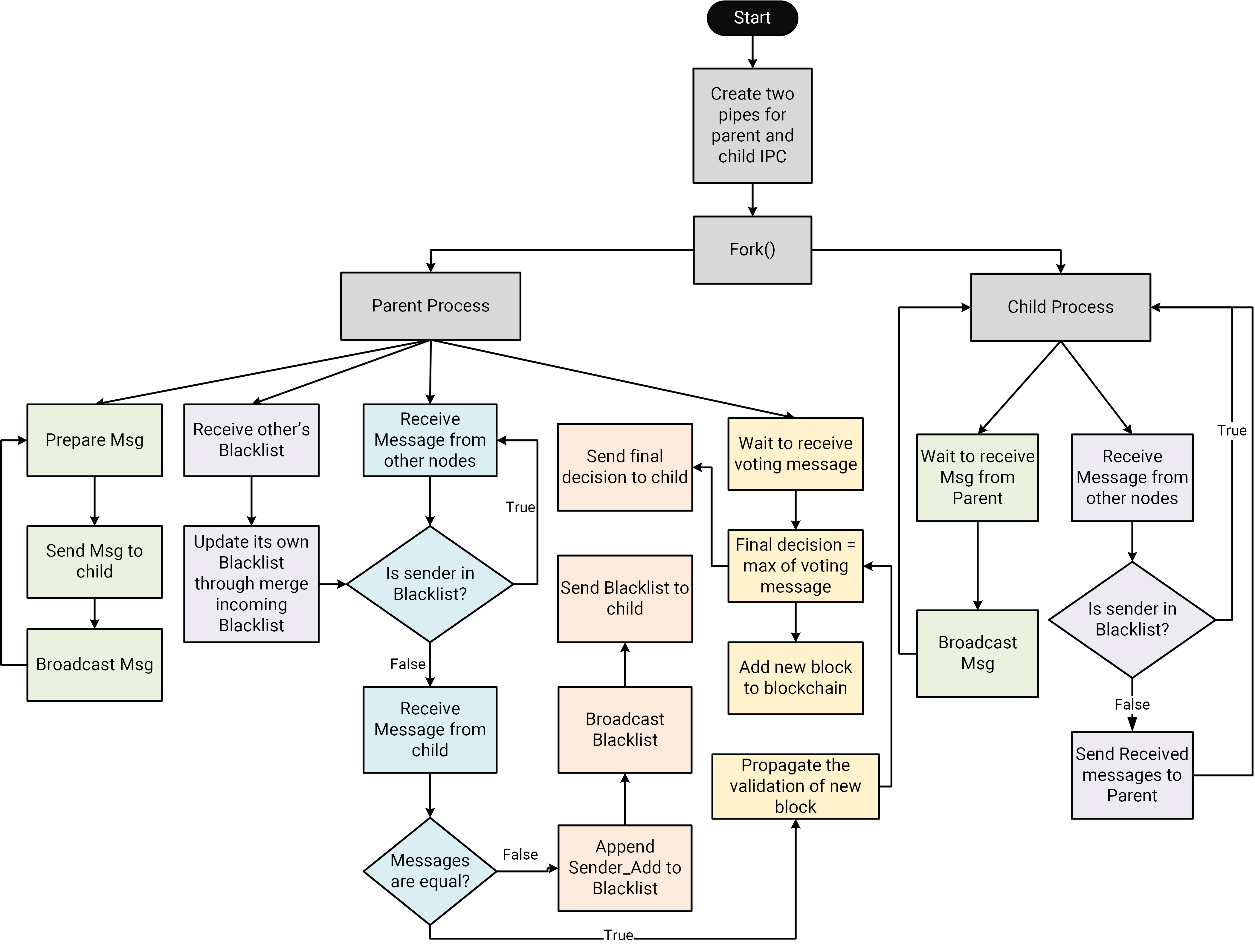}
    \caption{Two-Fold Byzantine Flowchart}
    \label{fig3}
\end{figure}
\subsection{TWO-FOLD BFT Algorithm}
In \hyperref[fig3]{Figure 3}, the flowchart provides a comprehensive overview of the operations conducted by both the parent and child processes within our system. The process begins with system initialization, during which two dedicated pipes for inter-process communication (IPC) are established to facilitate communication between the child and parent processes. This is followed by a fork operation to create a child process that runs parallel to the parent. The detailed operations of each forked process are described through the algorithm and illustrated in the flowchart.

In Parent Algorithm(\hyperref[algo1]{Algorithm 1}), upon receiving a message, the parent first checks if the sender is on the blacklist (lines 17-19). If so, it ignores the message and waits for another. If the sender is not blacklisted, the parent evaluates the message content. Should the message contain a blacklist update, the parent updates its own blacklist (lines 20-21). If the message pertains to a transaction, it is added to the inbox and retrieved from the child via the pipe. If both the parent and child receive the same transaction content and the sequence numbers match, the vote is deemed 'valid'; otherwise, the vote is 'not valid,' and the parent broadcasts its vote message (lines 22-31). Note that the vote message contains complete information about the specific transaction. In a situation where the transaction contents between parent and child are not equal, the sender is added to the blacklist, and the blacklist is then broadcast to all nodes excluding blacklist members (lines 32-35).

If the message involves voting, the parent counts the votes, and based on the highest tally, decides whether to add the transaction to the blockchain (lines 36-38).The 'prepare\_message()' function, used by each node, computes its transaction and prepares the message, including the sequence number, sender and receiver addresses (the latter excludes blacklisted members), and message type, which is a transaction (line 39). The definition of this function varies based on network functionality and may differ across networks. In line 40, both the MSG and the blacklist are sent to the child via the pipe, and in line 41, the MSG is sent to all nodes, excluding those on the blacklist.

Besides the parent algorithm, the child process (\hyperref[algo2]{Algorithm 2}) has relatively few tasks. It requires the pipes created by the parent, which it receives as input. During execution, the child process receives messages and a blacklist from the parent, and then broadcasts them to all nodes, excluding those on the blacklist (lines 1-3). In line 4, the child receives a message and checks whether it is on the blacklist. If it is, the child ignores it and waits for another message. If not, and if the message is a transaction, the child sends it to the parent via the pipe (lines 4-9).

\begin{figure}
    \centering
    \includegraphics[width=0.75\linewidth]{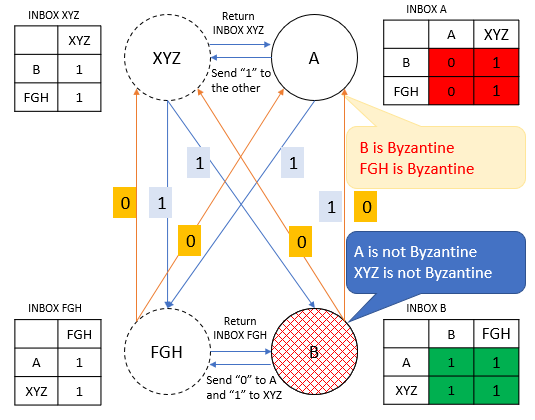}
    \caption{Schema of proposed algorithm}
    \color{black!100}
    \label{fig4}
\end{figure}

\hyperref[fig4]{Figure 4} illustrates the functioning of our system, which involves two nodes, A and B, with B being a Byzantine node. For simplicity, we consider that the messages exchanged are one-bit messages. Initially, no communication has been established between them, so they are unaware of each other's status. Each node creates its own child process and operates according to the algorithm previously described.
\begin{algorithm}
    \caption{Parent Algorithm}
    \begin{algorithmic}[1]
        \State Pipe\textsubscript{0} $\gets$ CreatePipe()   \Comment{Parent Pipe for IPC}
        \State Pipe\textsubscript{1} $\gets$ CreatePipe()   \Comment{Child Pipe for IPC}
        \State Process\textunderscore ID $\gets$ Fork()
        \If{Process\textunderscore ID is 1}
            \State Child(Pipe\textsubscript{1});
        \EndIf
        \If{Process\textunderscore ID is 0}
            \State Create a blank BlackList
            \State Create Inbox \Comment{Dynamic 2D Array with initial size of 1}
            \State Vote $\gets$ None
            \While{True}
                \State Message $\gets$ Receive\textunderscore Message() \Comment{All incoming messages from other nodes}
                \If{Message[Sender\textunderscore Add] is in BlackList}
                    \State Append Message[Sender\textunderscore Add] to BlackList
                    \State Broadcast Blacklist excluding Blacklist's Members
                    \State \textbf{Go to Line 11}
                \ElsIf{Message[Type] is "BlackList"}
                    \State Update BlackList with Message[Content]
                \ElsIf{Message[Type] is "Transaction"}
                    \State Inbox[Sender\textunderscore Add][0] $\gets$ Message[Content]
                    \State Inbox[Sender\textunderscore Add][1] $\gets$ Pipe\textsubscript{0}.Receive() \Comment{Message from sender will be stored in Inbox}
                    \If{Inbox[Sender\textunderscore Add][0] is Inbox[Sender\textunderscore Add][1]}
                        \State Vote $\gets$ "Valid"
                        \State Broadcast Vote excluding Blacklist's Members
                    \EndIf
                \ElsIf{Message[Type] is "Vote"}
                    \State Count Votes and Make Final Decision for Adding to Blockchain
                \EndIf
                \State MSG $\gets$ Prepare\textunderscore Message()
                \State Pipe\textsubscript{0}.Send(MSG, Blacklist) \Comment{Send MSG to Child}
                \State Broadcast MSG excluding Blacklist's Members
            \EndWhile
        \EndIf
        \label{algo1}
    \end{algorithmic}
\end{algorithm}
After the algorithm is run, the inboxes of nodes A and B are filled with messages received from each other and from their respective child processes. The comparison of these one-bit messages reveals that B is the Byzantine node due to the inconsistency in the messages received.
\begin{algorithm}
    \caption{Child Algorithm}
    \label{alg:cap}
    \begin{algorithmic}[1]
        \Require Pipe
        \While{True}
            \State MSG, Blacklist $\gets$ Pipe.Receive()
            \State Broadcast MSG excluding Blacklist's Members
            \State Message $\gets$ Receive\textunderscore Message() \Comment{All incoming messages from other nodes}
            \State Pipe.Send(Message) \Comment{Send Message to Parent}
        \EndWhile
        \label{algo2}
    \end{algorithmic}
\end{algorithm}
\section{Implementation and Verification}
As the algorithm describes, the final output of the system is a transaction sent by healthy nodes, which is added to the blockchain after voting. A simple comparison enables the distinction between healthy and Byzantine nodes. Messages received by the parent node are compared to those received by the child node. A Byzantine node is defined as one whose messages to a pair of parent and child nodes are not equal. Consequently, Byzantine nodes are separated from healthy ones and are stored on blacklists. The blacklist is shared among all healthy nodes, and ultimately, every healthy node will possess the same unique blacklist of Byzantine nodes. This mechanism ensures that all system processes can recognize whether a sender is Byzantine. There are two possible scenarios for detecting a Byzantine node using this method:
\begin{itemize}
    \item \textbf{The sender is not Byzantine:} if all its messages are identical.
    \item \textbf{The sender is Byzantine:} if it sends different messages to different processes. (It might also send messages arbitrarily, whether identical or different.)

\end{itemize}

If a node sends identical messages, it has not engaged in faulty behavior, and thus it is regarded as a healthy process. When nodes in a blockchain system are not aware of each other, they do not know whether each is Byzantine or not. Furthermore, each node views all other nodes as independent and does not know which node is under another’s control.

In the implementation, the content of the message is irrelevant to the functionality of the algorithm. However, for simplicity, our transaction messages consist of English letters.

A Byzantine node with adversarial intentions may send a 'healthy' message to our system, which results in no change or failure, thereby raising no concerns. To detect a Byzantine node, one of two conditions must be met: the messages are either identical or they are not. Thus, the probability that a parent process receives the same message as its child is determined by $p_{\text{equality}} = \frac{1}{2}$.

Nodes have a one-second probability to detect Byzantine nodes. However, if a Byzantine node sends identical messages to both the parent and child nodes, it cannot be detected as faulty.
While sending the same message might not indicate faulty behavior, it is considered a random occurrence affecting only certain parent-child pairs. Consequently, some nodes may detect the Byzantine node, while others might not. The overall probability of detecting Byzantine nodes is 1/2.     

As an example, let's consider a system with five processes: A, B, C, D, and E. Using their children's cooperation, each process receives and compares messages. The final comparison table is shown in \hyperref[table2]{Table 2}, where the letter 'e' stands for equality of child and parent messages, and 'd' indicates different messages from this pair. Process C is assumed to be both the sender of this message and a Byzantine process. According to the truth table, all possible states are displayed, and the probability of each state is as follows:
\[
P_{C \text{ is healthy}} = \left( p_{eA} \times \dots \times p_{\text{enth}} \right)
\]

\[
P_{eA}^{n-1} = \frac{1}{2^{n-1}}
\]

If the sender is healthy, it sends the same message to all processes. In \hyperref[table2]{Table 2}, this state is represented in the first row as <e, e, e, e>. States containing one or more 'd' letters indicate that Byzantine nodes sent different messages and engaged in malicious behavior. With this approach, if even one healthy node detects Byzantine behavior, it adds that node's identity to its blacklist and shares this with other healthy nodes. Eventually, all healthy nodes will be aware of the Byzantine node's identity.
\[P_{\text{C is Byzantine}} = 1- \frac{1}{2^{n-1}}\]

Among all the possible states for detecting a Byzantine node, just one state containing a 'd' is sufficient for detection. In this example, with five processes, four are healthy parents. Each of these parents is capable of detecting Byzantine nodes based on their shared blacklist, with a probability of:
\[P_{\text{C is Byzantine}} = 1- \frac{1}{2^{5-1}} = 1-\frac{1}{16}=0.9375\]
\begin{table}
  \centering
  \begin{tabular}{|c|c|c|c|c|c|c|c|c|} \hline 
       A&  A'&  B&  B'& C &  D&  D'&  E& E'\\ \hline 
       \multicolumn{2}{|c|}{e}&  \multicolumn{2}{|c|}{e}& \multirow{16}{*}{\rotatebox{90}{C is a Byzantine Node.}} &\multicolumn{2}{|c|}{e} &\multicolumn{2}{|c|}{e}\\
       \multicolumn{2}{|c|}{e}&  \multicolumn{2}{|c|}{e}&  &\multicolumn{2}{|c|}{e}&  \multicolumn{2}{|c|}{d}\\
       \multicolumn{2}{|c|}{e}&  \multicolumn{2}{|c|}{e}&  &\multicolumn{2}{|c|}{d}&  \multicolumn{2}{|c|}{e}\\
       \multicolumn{2}{|c|}{e}&  \multicolumn{2}{|c|}{e}&  &\multicolumn{2}{|c|}{d}&  \multicolumn{2}{|c|}{d}\\
       \multicolumn{2}{|c|}{e}&  \multicolumn{2}{|c|}{d}&  &\multicolumn{2}{|c|}{e}&  \multicolumn{2}{|c|}{e}\\
       \multicolumn{2}{|c|}{e}&  \multicolumn{2}{|c|}{d}&  &\multicolumn{2}{|c|}{e}&  \multicolumn{2}{|c|}{d}\\
       \multicolumn{2}{|c|}{e}&  \multicolumn{2}{|c|}{d}&  &\multicolumn{2}{|c|}{d}&  \multicolumn{2}{|c|}{e}\\
       \multicolumn{2}{|c|}{e}&  \multicolumn{2}{|c|}{d}&  &\multicolumn{2}{|c|}{d}&  \multicolumn{2}{|c|}{d}\\
       \multicolumn{2}{|c|}{d}&  \multicolumn{2}{|c|}{e}&  &\multicolumn{2}{|c|}{e}&  \multicolumn{2}{|c|}{e}\\
       \multicolumn{2}{|c|}{d}&  \multicolumn{2}{|c|}{e}&  &\multicolumn{2}{|c|}{e}&  \multicolumn{2}{|c|}{d}\\
       \multicolumn{2}{|c|}{d}&  \multicolumn{2}{|c|}{e}&  &\multicolumn{2}{|c|}{d}&  \multicolumn{2}{|c|}{e}\\
       \multicolumn{2}{|c|}{d}&  \multicolumn{2}{|c|}{e}&  &\multicolumn{2}{|c|}{d}&  \multicolumn{2}{|c|}{d}\\
       \multicolumn{2}{|c|}{d}&  \multicolumn{2}{|c|}{d}&  &\multicolumn{2}{|c|}{e}&  \multicolumn{2}{|c|}{e}\\
       \multicolumn{2}{|c|}{d}&  \multicolumn{2}{|c|}{d}&  &\multicolumn{2}{|c|}{e}&  \multicolumn{2}{|c|}{d}\\
       \multicolumn{2}{|c|}{d}&  \multicolumn{2}{|c|}{d}&  &\multicolumn{2}{|c|}{d}&  \multicolumn{2}{|c|}{e}\\
       \multicolumn{2}{|c|}{d}&  \multicolumn{2}{|c|}{d}&  &\multicolumn{2}{|c|}{d}&  \multicolumn{2}{|c|}{d}\\ \hline   
  \end{tabular}
  \caption{Truth Table for 5 nodes}
  \label{table2}
\end{table}

In our study, we conducted a simulation of the proposed algorithm using Python, supported by various libraries and Linux shell scripts. This section details the libraries used, their specific roles in our simulation, and discusses the simulation parameters, leading into a discussion of the results we achieved. One of the critical challenges was managing network configurations for forked processes, specifically differentiating the IP addresses of child processes from their parent processes. To address this, each parent-child pair, despite being on the same machine, was assigned distinct network namespaces. This separation was facilitated using Python’s os library to fork child processes and shell scripts to configure individual network namespaces and virtual Ethernet devices through the subprocess library.

For inter-process communication (IPC), we employed the os library's pipe mechanism, crucial for data transfer between parent and child processes. For more complex or asynchronous communication, we utilized libraries like ZeroMQ for its robust messaging capabilities. In the realm of blockchain simulation, we leveraged Python libraries such as PyChain for blockchain modeling, and pycrypto and hashlib for cryptographic functions, ensuring the integrity and security of our simulations.

Our simulation involved N parent nodes, each with a corresponding child node, totaling 2N nodes. We focused on parent nodes under the assumption that child nodes acted identically to their parents. We set a maximum iteration cap of 100, though our algorithm typically converged much sooner. We designed scenarios varying the number of Byzantine nodes present: for N=100, we considered t Byzantine nodes within \{30, 50, 70, 90\}, and for N=1000, within \{300, 500, 700, 900\}. These parameters facilitated a comparative analysis with existing algorithms, testing scenarios where traditional algorithms might falter.

The effectiveness of these algorithms was evaluated under the assumption they could identify Byzantine nodes at their maximum capacity, even if the actual number might exceed this threshold. We measured message complexity as O(n\textsuperscript{2}) and set the iteration interval for clearing the blacklist to every three iterations. This periodic reset serves two pivotal purposes: it allows for the reintegration of nodes previously classified as Byzantine, accommodating the dynamic nature of node behavior over time, and accounts for potential anomalies in network communication, such as delayed message delivery. This strategy of periodic reassessment is essential to maintain the accuracy and reliability of our node classification process, ensuring that our algorithm remains robust against both genuine behavioral changes and unforeseen network challenges.
\begin{figure}[ht]
    \centering
    \begin{minipage}{0.45\textwidth}
        \centering
        \includegraphics[width=\textwidth]{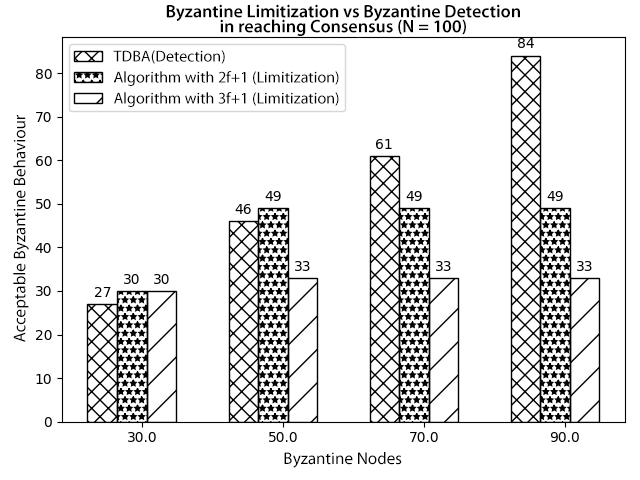}
        \caption{Comparison between Models (N = 100)}
        \label{fig5}
    \end{minipage}
    \hfill
    \begin{minipage}{0.45\textwidth}
        \centering
        \includegraphics[width=\textwidth]{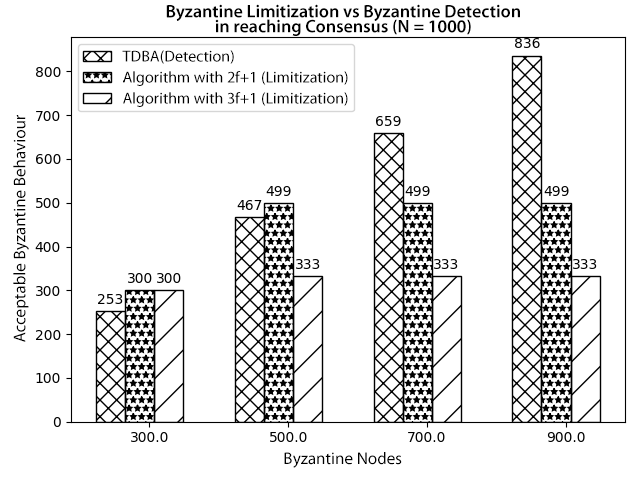}
        \caption{Comparison between Models (N=1000)}
        \label{fig6}
    \end{minipage}
\end{figure}
\section{Simulation Results}
In \hyperref[fig5]{Figure 5}, we explore scenarios where N=100, examining other algorithms' performance at full capacity. Despite theoretical capabilities, these algorithms can only tolerate a limited number of Byzantine nodes, set at t within \{30, 50, 70, 90\}. For lower counts (30 and 50), some algorithms managed to perform adequately. However, our algorithm outperformed others by detecting over 83\% of the Byzantine nodes. At higher counts (70 and 90), where other algorithms faltered, ours successfully identified more than 85\% of the nodes.

The second scenario, illustrated in \hyperref[fig6]{Figure 6} for N=1000, follows a similar pattern in a larger network. Here, t was set at \{300, 500, 700, 900\}. Our algorithm consistently detected over 89\% of Byzantine nodes at lower counts and over 92\% at higher counts, demonstrating superior detection capabilities and robustness.

\begin{figure}[h]
    \centering
    \includegraphics[width=0.63\textwidth]{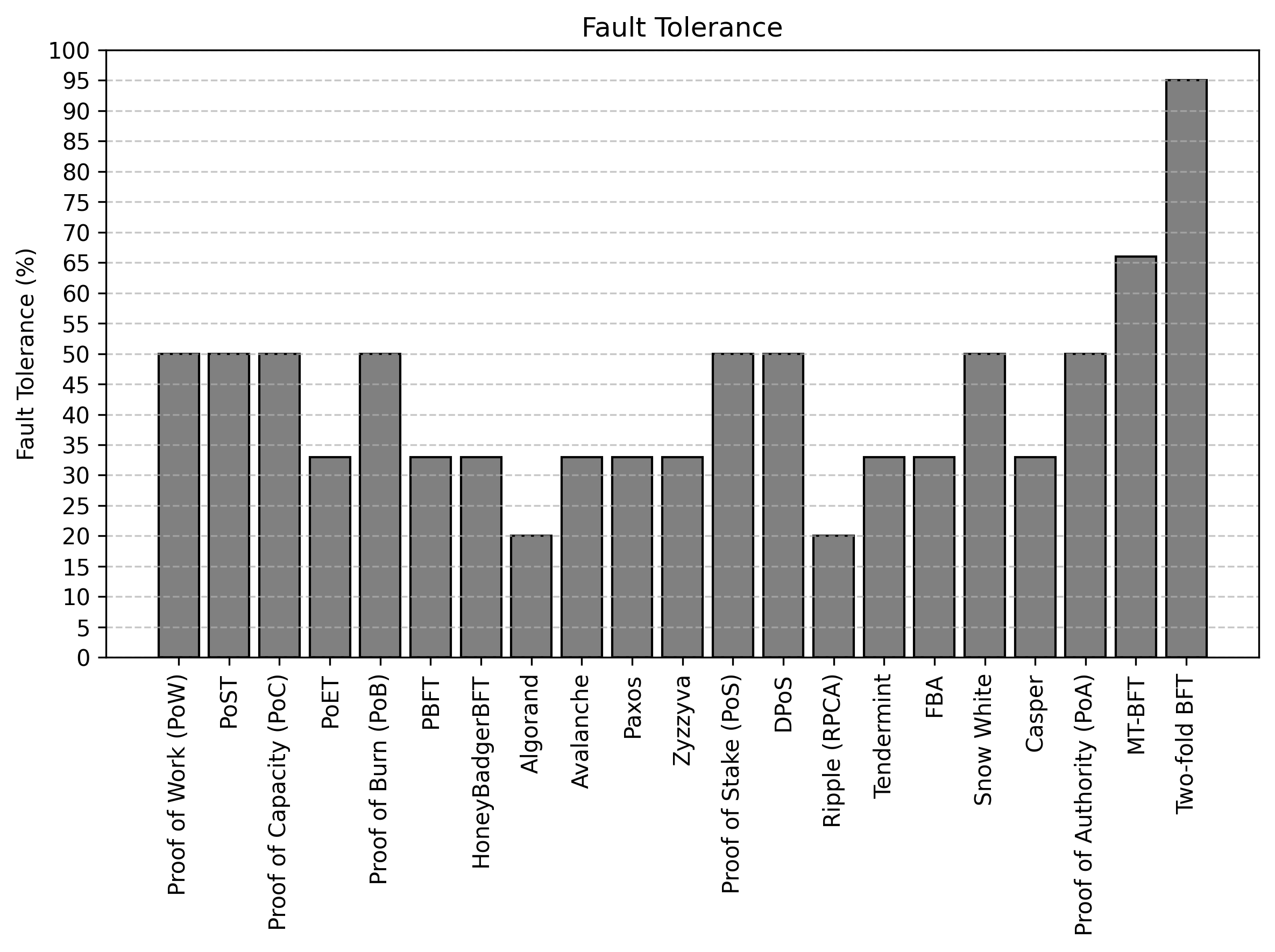}
    \caption{Fault tolerance (percentage) of consensus protocols}
    \label{fig7}
\end{figure}

Furthermore, in \hyperref[fig7]{Figure 7}, we compare our algorithm’s fault tolerance against others documented in the literature, observing that our method detects upwards of 95\% of Byzantine nodes—significantly higher than the 66\% maximum reported by Multi-Threshold Byzantine Fault Tolerance. Algorand and Ripple protocols tolerating only one-fifth of Byzantine nodes, while others like PBFT and Zyzzyva manage up to 33\% and the best up to 66\%. Our method can handle over 95\% of faults, identifying Byzantine behavior effectively even as node counts increase.

These percentages indicate how many nodes exhibit Byzantine behavior due to faults. As previously explained in our method outlined in section 4, it's possible that if all nodes are in just one state, they may receive different messages, implying Byzantine behavior across all nodes. However, using formula (5) and with 5 nodes, we can detect Byzantine nodes with 94\% accuracy. With a greater number of nodes, this accuracy could exceed 95\%.

\begin{figure}[ht]
    \centering
    \begin{subfigure}[b]{0.48\textwidth}
        \centering
        \includegraphics[width=\linewidth]{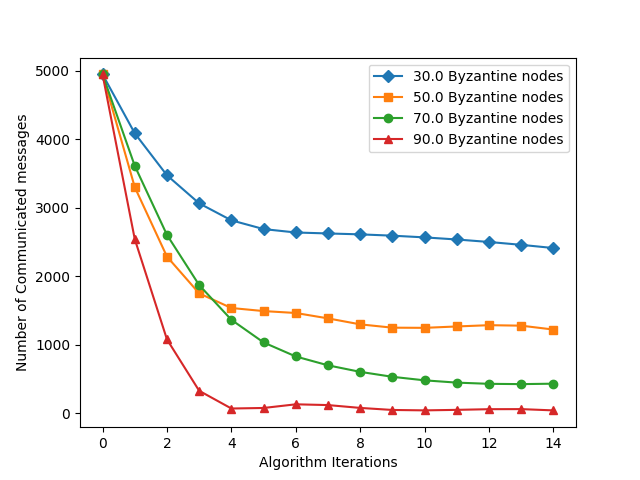}
        \caption{Without considering iteration intervals (N = 100)}
        \label{fig8a}
    \end{subfigure}
    \hfill
    \begin{subfigure}[b]{0.48\textwidth}
        \centering
        \includegraphics[width=\linewidth]{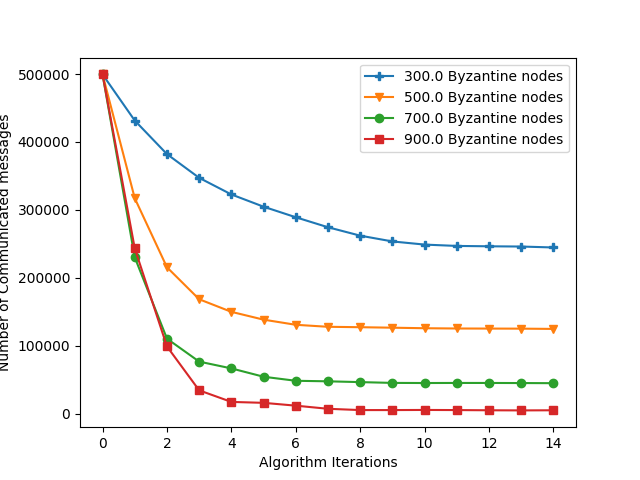}
        \caption{Without considering iteration intervals (N = 1000)}
        \label{fig8b}
    \end{subfigure}
    \hfill
    \begin{subfigure}[b]{0.48\textwidth}
        \centering
        \includegraphics[width=\linewidth]{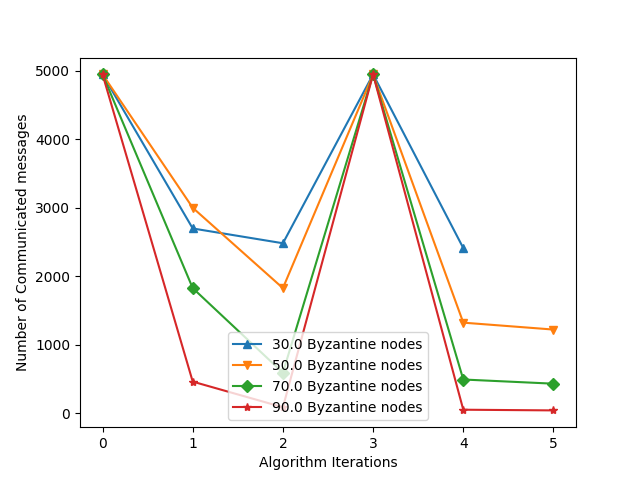}
        \caption{With considering iteration intervals (N = 100)}
        \label{fig8c}
    \end{subfigure}
    \hfill
    \begin{subfigure}[b]{0.48\textwidth}
        \centering
        \includegraphics[width=\linewidth]{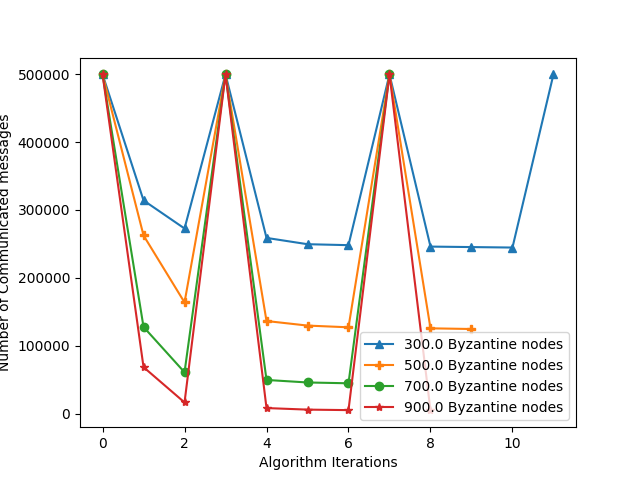}
        \caption{With considering iteration intervals (N = 1000)}
        \label{fig8d}
    \end{subfigure}
    \caption{Number of processed communicated messages}
\end{figure}

We also assessed computational complexity by monitoring message exchanges, which remained at O(n\textsuperscript{2}) despite strategies to reduce message complexity by excluding those to and from detected Byzantine nodes. \hyperref[fig8a]{Figure 8(a)} and \hyperref[fig8b]{Figure 8(b)} show our message complexity management, while \hyperref[fig8c]{Figure 8(c)} and \hyperref[fig8d]{Figure 8(d)} depict periodic reassessments allowing well-behaved nodes previously labeled as Byzantine to reintegrate, ensuring accurate long-term node classification.
\section{Conclusion}
This paper proposes an algorithm for Byzantine consensus through detecting the Byzantine nodes within a blockchain without constraining the number of Byzantine nodes. By eliminating constraints on the number of Byzantine nodes, the algorithm will be more robust and it will be practical in ideal systems. We presented the Tow-Fold BFT algorithm as a graph that had child and parent relations. This child process is a dependent process of its parent, sending what the parent process sends and delivering the received messages to its parent. A parent compares the messages it receives with those it receives from its child. When the parent finds a contradiction in the messages from the same sender, the sender is added to its blacklist. All healthy nodes share the blacklist. 

As a result, healthy nodes are aware of the Byzantine process because they share the same blacklist. Consequently, healthy nodes refrain from voting on messages from Byzantine nodes. This clearly indicates that the Two-Fold BFT algorithm can detect Byzantine nodes that behave arbitrarily and send malicious messages with a probability of more than 95\%. In our algorithm, fault tolerance entails isolating faulty nodes and exclusively engaging with healthy ones. This algorithm exhibits over 95\% fault tolerance for Byzantine nodes that exhibit Byzantine behavior. 

\section{Future Work}
Looking ahead, several promising avenues exist for extending the research presented in this paper. Firstly, future studies could aim to simplify the complexity of child processes while also reducing the number of messages exchanged. Additionally, the integration of more sophisticated reinforcement learning algorithms could enhance the accuracy and efficiency of our results, especially given the dynamic environments we have encountered. Another potential area for future work involves developing security algorithms for message communication between nodes in the network, an aspect not addressed in the current algorithm. Lastly, we envisage a mechanism that allows Byzantine nodes to auto-configure themselves and transition to well-behaved states autonomously.


\bibliographystyle{unsrt}  
\bibliography{references}

\end{document}